\documentstyle[12pt,titlepage,epsf]{article}
\def\_{\rule{.3em}{.15ex}} 
\newcommand{\scs}{\scriptscriptstyle}
\newcommand{\be}{\begin{equation}}
\newcommand{\ee}{\end{equation}}
\newcommand{\bea}{\begin{eqnarray}}
\newcommand{\eea}{\end{eqnarray}}
\newcommand{\f}{\frac}

\newcommand{\al}{\alpha_s}
\newcommand{\bsg}{$b \to s \gamma $ }

\newcommand{\cug}{$c \to u \gamma $ }

\begin{document}
\begin{titlepage}

 \begin{flushright}
  {\bf ZU-TH 6/96\\
       SLAC-PUB-7120\\
       hep-ph/9603417\\
       March 1996}
 \end{flushright}

 \begin{center}
  \vspace{0.6in}

\setlength {\baselineskip}{0.3in}
  {\bf \Large The $c \rightarrow u \gamma$ Contribution 
to Weak Radiative Charm Decay}
\vspace{0.7in} \\
\setlength {\baselineskip}{0.2in}

{\large  Christoph Greub$^{^{1,2}}$ \\}
{\it Stanford Linear Accelerator Center\\
Stanford University, Stanford, California 94309, USA}
\vspace{0.3in}

{\large Tobias Hurth$^{^{1,3}}$, 
        \ Miko{\l}aj Misiak$^{^{1,4}}$ 
          \ and \ Daniel Wyler$^{^1}$ \\}
{\it Institut f\"ur Theoretische Physik der Universit\"at Z\"urich,\\
Winterthurerstrasse 190, 8057 Z\"urich, Switzerland}.

\vspace{1in} 
{\bf Abstract \\} 
\end{center} 
\setlength{\baselineskip}{0.3in} 

The $c \rightarrow u \gamma$ transition does not occur at the tree
level in the Standard Model and is strongly GIM-suppressed at one
loop.  The leading QCD logarithms are known to enhance the amplitude
by more than an order of magnitude. We point out that the amplitude
increases further by two orders of magnitude after including the
complete 2-loop QCD contributions. Nevertheless, $\Delta S =0$
radiative decays of charmed hadrons remain dominated by the \linebreak
$c \rightarrow d \bar{d} u \gamma$ and $c \rightarrow s \bar{s} u \gamma$
quark subprocesses.  \vspace{0.5in}

\vspace{0.5in}

\setlength {\baselineskip}{0.2in}
\noindent \underline{\hspace{2in}}\\ 
$^1$ {\footnotesize Work supported in part by Schweizerischer
Nationalfonds.}\\
$^2$ {\footnotesize Work supported in part by the Department of
Energy, contract DE-AC03-76SF00515.}\\
$^3$ {\footnotesize Address after March 1996: SUNY at Stony Brook, 
Stony Brook NY 11794-3840, USA.}\\
$^4$ {\footnotesize Partially supported by the Committee for
Scientific Research, Poland.}

\end{titlepage}

\setlength {\baselineskip}{0.3in}

\noindent {\bf 1.} Inclusive heavy flavor decay can be systematically
analyzed with help of an expansion in inverse powers of the heavy
quark mass \cite{CGG}. At the leading order in such an expansion, the
inclusive hadronic decay rate is given by the perturbatively
calculable free quark decay rate. For charmed hadron decays, this
procedure can certainly be questioned since the charm quark mass is
not really much larger than the QCD scale $\Lambda$. Nevertheless,
some properties of charmed hadrons have been analyzed succesfully with
help of HQET \cite{N}. Therefore, it is also of interest to calculate
inclusive decay rates of these particles at the leading order in the
heavy mass expansion.

        In the present letter, we consider weak radiative charm decay
\cite{BGHP,BFO}. For definiteness, we restrict ourselves to $\Delta
S=0$ processes. At the leading order in electroweak and strong
interactions, there are three contributions to the $\Delta S=0$ charm
quark radiative decay: \cug, $c \to u d \bar{d} \gamma$ and $c \to u s
\bar{s} \gamma$. At higher orders in the strong interactions, there
are virtual corrections to these decays as well as new decay modes
with more gluons and quark-antiquark pairs in the final state.

        In the absence of QCD, the \cug amplitude is enormously
suppressed by GIM cancellations and by small Cabibbo-Kobayashi-Maskawa
(CKM) matrix elements. The leading logarithmic contribution turns out
to be considerably larger, but is still afflicted with a small CKM
coefficient. This raises a hope that higher order terms can be the
most important ones, if they are not suppressed by small CKM factors.

        In this paper, we show that the \cug process is completely
dominated by a two-loop term which has not been considered so far in
perturbative analyses of charm decays. Furthermore, we point out that
decays with more gluons which are also of higher order in the strong
interactions have comparable amplitudes.

        The \cug channel cannot be separated kinematically from the
accompanying modes by making a cut on the photon energy spectrum,
because all the decay products are light. This is contrary to the \bsg
decay which can be separated from the $b \to s c \bar{u} \gamma$ and
other similar channels by selecting photon energies above the charm
production threshold. In the \cug case, we might still hope to be able
to separate this channel by looking at the shape of photon energy
spectrum. However, the possibility of doing so depends crucially on
how large the \cug amplitude is.

\newpage
\noindent {\bf 2.} We begin with the \cug process and briefly review
the known contributions which come at the (formally) lower orders of
perturbation theory.  All momentum invariants in a two-body decay are
expressible in terms of masses. Therefore, we can represent the
amplitude by a tree diagram with a single local vertex. For the \cug
decay in the Standard Model, the relevant local interaction reads
\be \label{Lint}
L_{int} = -\f{4 G_{\scs F}}{\sqrt{2}} \; A \; \f{e}{16 \pi^2} \; m_c 
                (\bar{u} \sigma_{\mu\nu} P_{\scs R} c) F^{\mu\nu},
\ee 
where $P_{\scs R} = (1+\gamma_5)/2$ and we have neglected the
$u$-quark mass. The coefficient $A$ nontrivially depends on light
quark masses (see below). Thus, it should not be misinterpreted as a
Wilson coefficient.

\begin{figure}[htb] 
\epsfysize = 0.9in
\epsffile{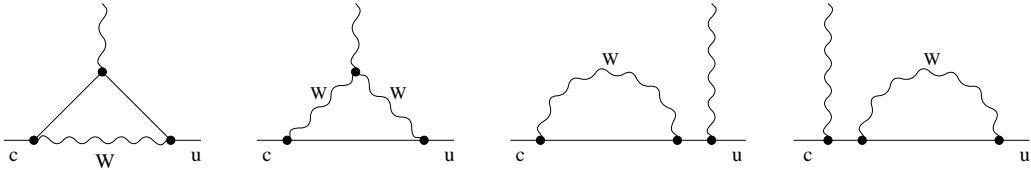}
\caption{One-loop diagrams generating the \cug transition.}
\label{fig.1loop}
\end{figure}

        As any flavor changing neutral current process, the \cug
amplitude arises  in the Standard Model only at the one-loop level. The
relevant diagrams are shown in fig. \ref{fig.1loop} and give the
following contribution to the coefficient $A$:
\be \label{1loop}
\Delta A_{\rm 1\;loop} \simeq -\f{5}{24} 
                \sum_{q=d,s,b} V^*_{cq} V_{uq} \left( \f{m_q}{M_W} \right)^2.
\ee
The CKM factors in the above equation have very different orders of
magnitude
\be
|V^*_{cd} V_{ud}| \simeq |V^*_{cs} V_{us}| \simeq 0.22
\hspace{1cm} \mbox{and} \hspace{1cm}
|V^*_{cb} V_{ub}| \simeq (1.3 \pm 0.4) \times 10^{-4}.
\ee
Consequently, $|\Delta A_{\scs 1\;loop}| \sim 2 \times 10^{-7}$. 
The extraordinary smallness of this number is due to the
tiny factors $(m_q/M_W)^2$ for the light quarks and to the small CKM
angles in the $b$-quark contribution.

        Since the important suppression factors are independent of
gauge couplings, it is possible that higher orders in perturbation
theory give dominant contributions to the radiative amplitude
considered because they may not suffer the same dramatic supressions
and are reduced only by powers of the gauge couplings.

        A natural first attempt to include higher orders is to resum
short distance QCD corrections in the leading-logarithmic
approximation, by analogy to the \bsg decay where they bring sizeable
enhancements \cite{BMMP}.  In order to systematically include these
contributions, one introduces two effective hamiltonians
\bea
H_{eff}(M_W > \mu > m_b) &=& \f{4 G_{\scs F}}{\sqrt{2}} \sum_{q=d,s,b} 
V^*_{cq} V_{uq} [ C_1(\mu) O^q_1 + C_2(\mu) O^q_2 ] \label{H1}\\
H_{eff}(m_b > \mu > m_c) &=& \f{4 G_{\scs F}}{\sqrt{2}} \sum_{q=d,s} 
V^*_{cq} V_{uq} [ C_1(\mu) O^q_1 + C_2(\mu) O^q_2 + 
                                 \sum_{i=3}^8 C_i(\mu) O_i ] \label{H2},
\eea
where
\bea
O^q_1 &=& (\bar{u}_{\alpha} \gamma_{\mu} P_{\scs L} q_{\beta})
          (\bar{q}_{\alpha} \gamma^{\mu} P_{\scs L} c_{\beta}), 
                                                        \hspace{1cm} q=d,s,b\\
O^q_2 &=& (\bar{u}_{\alpha} \gamma_{\mu} P_{\scs L} q_{\alpha}) 
          (\bar{q}_{\beta}  \gamma^{\mu} P_{\scs L} c_{\beta}), 
                                                        \hspace{1cm} q=d,s,b\\
O_7   &=&  \f{e}{16 \pi^2} \; m_c
      (\bar{u}_{\alpha} \sigma_{\mu\nu} P_{\scs R} c_{\alpha}) F^{\mu\nu}.
\eea
The remaining operators $O_i$ are given explicitly in eqn.~(18) of
ref.~\cite{BGHP}. We do not present them here because they are
inessential for our discussion.

        As indicated in eqns.~(\ref{H1}) and (\ref{H2}), each of the
hamiltonians is valid in a different range of the renormalization
scale $\mu$. Now, we neglect all terms which are suppressed by
additional powers of $1/M^2_W$ such as the ones in eqn.~(\ref{1loop}).
Due to CKM unitarity, the operators $O_3$,...,$O_8$ are not generated
by QCD renormalization at scales $\mu > m_b$.

        Finding the coefficients $C_i(\mu=M_W)$ and performing the
renormalization group evolution from $\mu=M_W$ to $\mu=m_c$ is by now
standard \cite{GW}. Analogously to the \bsg analysis of
ref.~\cite{BMMP}, we introduce an ``effective'' anomalous dimension
matrix. It reads
\bea
\hat{\gamma}^{(0)eff} = \left[
\begin{array}{cccccccc}
\vspace{0.2cm}
-2 & 6 &     0     &      0     &     0    &       0     &              0                &        3       \\
\vspace{0.2cm}
 6 &-2 &-\f{2}{9}  &  \f{2}{3}  &-\f{2}{9} &  \f{2}{3}   &    8 Q_1 + \f{16}{27} Q_2     &   \f{70}{27}   \\
\vspace{0.2cm}
 0 & 0 &-\f{22}{9} & \f{22}{3}  &-\f{4}{9} &  \f{4}{3}   &        \f{464}{27} Q_2        & \f{140}{27}+3f \\
\vspace{0.2cm}
 0 & 0 &6-\f{2}{9}f&-2+\f{2}{3}f&-\f{2}{9}f&  \f{2}{3}f  &  8 \overline{Q} + \f{16}{27} f Q_2 &  6+\f{70}{27}f \\
\vspace{0.2cm}
 0 & 0 &     0     &      0     &     2    &      -6     &         -\f{32}{3} Q_2        &  -\f{14}{3}-3f \\
\vspace{0.2cm}
 0 & 0 &-\f{2}{9}f & \f{2}{3}f  &-\f{2}{9}f&-16+\f{2}{3}f& -8 \overline{Q} + \f{16}{27} f Q_2 & -4-\f{119}{27}f\\
\vspace{0.2cm}
 0 & 0 &     0     &      0     &     0    &       0     &           \f{32}{3}           &        0       \\
\vspace{0.2cm}
 0 & 0 &     0     &      0     &     0    &       0     &         \f{32}{3} Q_2         &    \f{28}{3}   \\
\end{array} \right] \nonumber \eea
        For the renormalization group evolution from $\mu = m_b$ to
$\mu = m_c$ in the \cug case, one needs to substitute $f=4$, $\;Q_1 =
Q_d = -\f{1}{3}$, $\;Q_2 = Q_u = \f{2}{3}$ and $\overline{Q} = 2 Q_u +
2 Q_d = \f{2}{3}$.

        For the evolution from $\mu = M_W$ to $\mu = m_b$ in the \bsg
case, one would need to substitute $f=5$, $\;Q_1 = Q_u$, $\;Q_2 =
Q_d$ and $\overline{Q} = 2 Q_u + 3 Q_d = \f{1}{3}$. In this case, the
matrix $\hat{\gamma}^{(0)eff}$ given in Appendix A of ref.~\cite{BMMP}
would be reproduced.

The resulting leading logarithmic contribution to $A$ in
eqn.~(\ref{Lint}) reads
\be \label{LLA}
\Delta A_{\scs LLA} = -V^*_{cb} V_{ub} \; C_7^{eff}(m_c),
\ee
where
\be \label{c7eff}
C_7^{eff}(m_c) = \sum_{i=1}^8 
\left( \f{\al(m_b)}{\al(m_c)} \right)^{z_i} 
\left[ x_i C_1(m_b) + y_i C_2(m_b) \right].
\ee
The coefficients $C_{1,2}(m_b)$ are given by
\be
C_{1,2}(m_b) = \f{1}{2} \left( \f{\al(M_W)}{\al(m_b)} \right)^{\f{6}{23}} 
           \mp \f{1}{2} \left( \f{\al(M_W)}{\al(m_b)} \right)^{-\f{12}{23}},
\ee
and the numbers $x_i$, $y_i$ and $z_i$ read
\be \label{numbers}
\begin{array}{ccccccccc}
\vspace{0.1cm}
x_i = (& -\f{65710}{18413},  & \f{22173}{8590}, &  \f{2}{5}, &    0, 
       &      0.6524,        &     -0.0532,     &  -0.0034,  & -0.0084 ),\\
\vspace{0.1cm}
y_i = (& -\f{675158}{165717},& \f{23903}{8590}, &  \f{2}{5}, &    0, 
       &      0.8461,        &      0.0444,     &   0.0068,  & -0.0059 ),\\
z_i = (&     \f{14}{25},     &    \f{16}{25},   & \f{6}{25}, & -\f{12}{25}, 
       &      0.3469,        &      -0.4201,    &  -0.8451,  &   0.1317   ).
\end{array}
\ee
Note that the r.h.s. of eqn.~(\ref{c7eff}) vanishes in the formal
limit $\al(m_c) \to \al(m_b)$, as it should.

        Taking $m_b = 5\;{\rm GeV}$, $m_c = 1.5\;{\rm GeV}$ and
$\al(M_Z) = 0.12$ (which implies $\al(m_b)/\al(m_c) \simeq 0.67$), we
obtain
\be \label{numLLA} 
|\Delta A_{\scs LLA}| = 
[ 0.001 C_1(m_b) + 0.055 C_2(m_b)]\;|V^*_{cb} V_{ub}| = 
0.060 \; |V^*_{cb} V_{ub}| \simeq (8 \pm 3) \times 10^{-6}.
\ee
This result is more than an order of magnitude larger than the (formally)
leading order one in eqn.~(\ref{1loop}): Including logarithmic QCD
contributions replaces the powerlike GIM suppression factors
$(m_q/M_W)^2$ by a logarithmic function of $m_b/m_c$.

        This surprising enhancement was noted some time ago and
discussed in detail in ref. \cite{BGHP}. However, our coefficient
$C_7^{eff}(m_c)$ in eqn.~(\ref{c7eff}) is 5 times smaller than the one
obtained there.  This difference arises mainly because the operators
$O_1^b$ and $O_2^b$ have not been taken into account in
ref. \cite{BGHP}. Their presence above $\mu = m_b$ is essential for
cancellation of additive logarithmic QCD contributions at these
renormalization scales.\vspace{0.2in}

{\bf 3.} Although the leading logarithmic result (\ref{numLLA}) is
much larger than the purely electroweak contribution (\ref{1loop}),
there remains a strong cancellation between $s$ and $d$ loops whose
CKM factors are very similar in magnitude but have opposite
signs. Consequently, a suppression by $V^*_{cb} V_{ub}$ is still
present in eqn.~(\ref{numLLA}). 

\begin{figure}[htb] 
\epsfysize = 0.9in
\epsffile{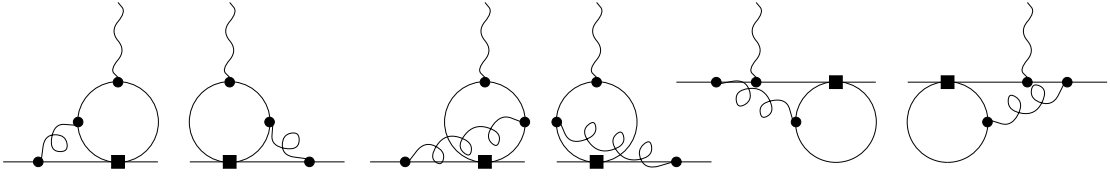}
\caption{Diagrams contributing to the two-loop \cug matrix element of
$O^q_2$.}
\label{fig.2loop}
\end{figure}

        This cancellation is circumvented when the functional
dependence of the amplitude on the $s$- and $d$-quark masses becomes
substantial. It turns out that this happens in the two-loop matrix
element of the effective hamiltonian in eqn.~(\ref{H2}). It is given
by the diagrams in fig. \ref{fig.2loop} with $O^s_2$ and $O^d_2$
operator insertions.\footnote{ 
Two-loop diagrams with $O^q_1$ insertions vanish due to their color
structure. One-loop \cug diagrams with $O^q_1$ or $O^q_2$ insertions
vanish for on-shell photons.}

After using unitarity of the CKM matrix, we obtain the following
contribution to the coefficient A:
\be \label{2loop}
A = V^*_{cs} V_{us} \f{\al(m_c)}{4 \pi} \; C_2(m_c) \;
         \{ f[(m_s/m_c)^2] - f[(m_d/m_c)^2] \} 
\hspace{0.4cm} + \hspace{0.4cm} {\cal O}(V^*_{cb} V_{ub}),
\ee
where
\be
C_2(m_c) = \f{1}{2} \left( \f{\al(M_W)}{\al(m_b)} \right)^{\f{6}{23}} 
                    \left( \f{\al(m_b)}{\al(m_c)} \right)^{\f{6}{25}} 
         + \f{1}{2} \left( \f{\al(M_W)}{\al(m_b)} \right)^{-\f{12}{23}}
                    \left( \f{\al(m_b)}{\al(m_c)} \right)^{-\f{12}{25}}.
\ee
The function $f$ can be extracted from an analogous computation
performed recently for the $b$-quark decay \cite{GHW}. It reads
\bea
f(z)  &=& -\frac{1}{243} \, \left\{ 576 \pi^2 z^{3/2} \right. 
\nonumber \\
&& \hspace{1.2cm}
+ \left[ 3672 -288 \pi^2 -1296 \zeta (3) + (1944-324 \pi^2) L +
108 L^2 + 36 L^3 \right] \, z 
\nonumber \\
&& \hspace{1.2cm}
+ \left[ 324 - 576 \pi^2 + (1728 - 216 \pi^2) L + 324 L^2 + 
36 L^3 \right] \, z^2
\nonumber \\
&& \hspace{1.2cm}        \left.                 +
\left[ 1296 - 12 \pi^2 + 1776 L - 2052 L^2 \right] \, z^3 \, \right\}
\nonumber \\
&& - \frac{4 \pi i}{81} \, \left\{
\left[ 144 -6 \pi^2 + 18 L + 18 L^2 \right] \, z \right. 
\nonumber \\
&& \hspace{1.2cm} + \left[ -54 -6 \pi^2 + 108 L + 18 L^2 \right] \, z^2 
\nonumber \\
&& \hspace{1.2cm} + \left.
\left[ 116 - 96 L  \right] \, z^3 \, \right\} 
\hspace{0.3cm} + \hspace{0.3cm} {\cal O}(z^4 L^4),
\eea
where $L = \log z$. This function is renormalization scheme
independent. All scheme dependent terms in the two-loop matrix
elements of $O^s_2$ and $O^d_2$ undergo GIM cancellation and only
affect the \ ${\cal O}(V^*_{cb} V_{ub})$ part of eqn.~(\ref{2loop}).

        In the interesting range $0.005 < z < 0.1$, the function
$f(z)$ is approximated (to 15\% accuracy) by
\be
f(z) \simeq  -19 z ( 1 - 2 z ) - 3 i z \log^2 z.
\ee

        Using the s-quark mass $m_s(\mu=m_c) = (0.17 \pm 0.03)\; {\rm
GeV}$ \cite{JM} (but keeping $m_c = 1.5\;{\rm GeV}$ fixed), we find
\be \label{num2loop}
|A| = |V^*_{cs} V_{us}| \f{\al(m_c)}{4 \pi} \; (0.86 \pm 0.19)
   = (4.7 \pm 1.0) \times 10^{-3}.
\ee
Thus, the two-loop amplitude is more than two orders of magnitude
larger than the leading logarithmic result in eqn.~(\ref{numLLA}) and
four orders of magnitude larger than the one-loop contribution in
eqn.~(\ref{1loop}). The three contributions to the coefficient $A$ are
summarized in table~1.

\begin{center}
\begin{tabular}{|c|c|}
\hline
One-loop electroweak diagrams & 
         $|\Delta A_{\scs 1\;loop}| \sim 2 \times 10^{-7}$\\
\hline
Leading logarithmic approximation & 
        $|\Delta A_{\scs LLA}| =  (8 \pm 3) \times 10^{-6}$\\
\hline
Dominant two-loop diagrams &
        $|A| = (4.7 \pm 1.0) \times 10^{-3}$\\
\hline
\end{tabular}\\
\vspace{0.2cm}
Table 1. Summary of the contributions to the coefficient $A$.
\end{center}

        The form of eqn. (\ref{2loop}) ensures us that no furher
enhancement of the perturbative amplitude is expected at even higher
orders. Any $\Delta S = 0$ charm decay must be Cabibbo-suppressed. The
suppression by $m_s/m_c$ must also remain. The latter suppression is
actually rather mild, as one can see from eqns. (\ref{cons}) and
(\ref{curr}) below. We have thus exhausted the possibility of finding
large contributions by considering higher orders in perturbation
theory.\vspace{0.2in}

\noindent {\bf 4.} The dominant two-loop contribution to the \cug
amplitude is suppressed by $(m_s/m_c)^2$. Since $m_s$ is smaller
(though not by much) than the $\bar{\Lambda}$ parameter of HQET, the
nonperturbative contribution to a D-hadron decay may be equally (or
more) important than the \cug contribution.

        Some part of the nonperturbative contribution may be taken
into accont by replacing current quark masses in the r.h.s. of
eqn.~(\ref{2loop}) by constituent ones. Since the ratio of constituent
masses is closer to unity than the ratio of current masses, one might
worry that nonperturbative contributions may tend to bring back the
GIM cancellations between $s$ and $d$ quarks. It is easy to explicitly
check that this is not the case. For the constituent masses $m_d^{con} =
0.3 \; {\rm GeV}$ and $m_s^{con} = 0.45 \; {\rm GeV}$ one obtains
\be \label{cons}
f[(m_s^{con}/m_c)^2] - f[(m_d^{con}/m_c)^2] \simeq -0.68 - 0.48\;i,  
\ee
while for the current masses one has
\be \label{curr}
f[(m_s/m_c)^2] - f[(m_d/m_c)^2] \simeq -0.24 - 0.68\;i.
\ee
We see that the resulting amplitude does not decrease (it is even
larger) and conclude that the enhancement effect is rather robust.

        Ignoring the nonperturbative terms and normalizing the \cug
rate to the semileptonic decay rate (analogously to what one usually
does in the \bsg case), we would obtain the following \cug
contribution to the branching ratio of $\Delta S = 0$ weak radiative
D-meson decay:
\be \label{BR}
\Delta BR[
\;D \stackrel{\scs \Delta S=0}{\longrightarrow} X \gamma]_{c \to u \gamma} 
= \f{\alpha_{\scs QED}}{\pi} \f{6 |A|^2}{|V_{cd}|^2} 
\; BR[\;D  \stackrel{\scs \Delta S=0}{\longrightarrow} X \bar{e} \nu].  
\ee
This amounts to roughly $5 \times 10^{-8}$ for $D^+$ and to $2.5
\times 10^{-8}$ for $D^0$.  Although it is dramatically increased over
the previous results, it remains very small.  Compared to $c \to u d
\bar{d} \gamma$ and $c \to u s \bar{s} \gamma$, the \cug channel is
suppressed (in branching ratio) by a factor of order $|A/V_{cd}|^2
\sim 5 \times 10^{-4}$. We conclude that there is little chance to
extract the \cug contribution even with use of the photon energy
spectra.

        The obtained branching ratio is roughly confirmed when one
looks at the exclusive decays $D \to \rho \gamma$.  Assuming its $BR$
to be down by about a factor of ten from the inclusive one, the
contribution of \cug amounts to an exclusive $BR$ of about $5 \times
10^{-9}$ while the four-Fermi processes would give $10^{-5}$.  This is
in satisfactory agreement with a recent sum rule calculation
\cite{KSW} of $D \to \rho \gamma$.

        Since the \cug contribution in eqn.~(\ref{BR}) turns out to be
suppressed by both $(\al(m_c)/\pi)^2$ and powers of $m_s/m_c$, it is
by no means the largest even among subdominant contributions. Decay
modes with more gluons may have larger $BR$.  For instance, the $c \to
u \gamma\; gluon$ contribution to $BR[\;D\stackrel{\scs\Delta
S=0}{\longrightarrow}X\gamma]$ is proportional only to linear
$\al(m_c)/\pi$. It can be written in the same form as eqn.~(\ref{BR})
but with $|A|$ replaced by $1.1 \times 10^{-2}$.  Consequently, it is
a factor of 5 larger in the rate than the \cug contribution. The
quoted numerical value of the $c \to u \gamma\; gluon$ contribution
has been obtained with use of analogous $b \to s \gamma\; gluon$
results from ref. \cite{AG} where appropriate replacements of electric
charges had to be made.\vspace{0.2in}

\noindent{\bf 5.} To conclude, we have pointed out an interesting
enhancement of the (formally) higher order contributions in the \cug
decay. Nevertheless, the corresponding contribution to the inclusive
weak radiative charmed hadron decay most probably remains screened by
other channels.\vspace{0.2in}

\noindent{\bf 6.} Note added: The present version of this paper differs
from the original one by correcting a mistake in the negligible
one-loop contribution in eqn.~(\ref{1loop}). It was noticed by
Q.~Ho-Kim and X.Y.~Pham in hep-ph/9906235.\vspace{0.2in}

\setlength {\baselineskip}{0.2in}
 
\end{document}